\documentclass[floatfix,preprint, onecolumn, prd, showpacs,nofootinbib]{revtex4-1}
\usepackage{natbib}
\usepackage{times}
\usepackage{amssymb,amsbsy,amsmath,amsfonts}
\usepackage{graphicx}
\usepackage{float}
\usepackage{morefloats}
\usepackage{rotating}
\usepackage{srcltx}
\usepackage{slashed}
\begin{document}

\title{Off-shell effects on the interaction of Nambu-Goldstone bosons and $D$ mesons}

\author{ M. Altenbuchinger$^1$ and Li-Sheng Geng$^{2,1}$  }
\affiliation{
$^1$Physik Department, Technische Universit\" at M\"unchen, D-85747 Garching, Germany\\
$^2$School of Physics and Nuclear Energy Engineering  and International Research Center for
Particles and Nuclei in the Cosmos, Beihang University,  Beijing 100191,  China}

 \begin{abstract}
 The Bethe-Salpeter equation in unitarized chiral perturbation theory is usually solved with the so-called on-shell approximation. The underlying argument is
 that the off-shell effects can be absorbed by the corresponding coupling constants and physical masses, which has been
 corroborated by the success of unitarized chiral perturbation theory in describing a variety of physical phenomena. Such an approximation needs to be scrutinized when applied to study the light-quark mass evolution of physical observables, as routinely performed nowadays. In the present work, we  propose to solve the Bethe-Salpeter equation with the
full off-shell terms of the chiral potentials and apply this formalism to
the description of the latest $n_f=2+1$  lattice QCD (LQCD) data on the scattering lengths of  Nambu-Goldstone bosons off  $D$ mesons.  It is shown that
 the LQCD data can be better described in this formalism than in the widely used on-shell approximation.
 On the other hand,  no qualitative difference between the on-shell and off-shell approaches is observed for the light-quark mass evolution of the scattering lengths,
 given the limited LQCD data and their relatively large uncertainties. We also show that the light-quark mass dependence of the $D^*_{s0}(2317)$ remains essentially the same in
 both approaches.
\end{abstract}


\date{\today}

\maketitle

\section{Introduction}
In the past two decades,  a lot of progress has been made
in applying nonperturbative approaches based on principles of effective field theories to
understand low-energy strong interaction phenomena. One prominent example is the combination of chiral Lagrangians with unitarization techniques, i.e., the  so-called unitarized chiral perturbation theory (UChPT)~
\cite{ Kaiser:1995eg,Dobado:1996ps,Oller:1997ti,Oset:1997it,Oller:1998hw,Kaiser:1998fi,Oller:1998zr,Oller:2000fj,Lutz:2001yb}. Compared to other phenomenological methods, UChPT has a more transparent link to the theory underlying the strong interactions, quantum chromodynamics (QCD), and, as a result, is in principle improvable in a systematic manner.  Over the years, it has provided new insights into  the nature of various hadrons,  from the well-established hadrons, such as the $\Lambda(1405)$ or the $N^*(1535)$~\cite{ Kaiser:1995eg,Oset:1997it}, to those of the newly observed  $XYZ$ particles, such as the $X(2175)$~\cite{MartinezTorres:2008gy} or the $X(3872)$~\cite{Wang:2013kva}.

At the heart of UChPT are the interaction kernels provided by chiral Lagrangians, which are
constrained by QCD and its approximate symmetries, such as chiral symmetry and heavy quark spin/flavor symmetry and their breaking pattern.  Exact two-body ($s$-channel) unitarity can be implemented in different ways and a widely adopted approach is the Bethe-Salpeter (BS) approach~\cite{Salpeter:1951sz}. It is well known that the BS approach provides exact unitarity but loses crossing symmetry, compared to
conventional chiral perturbation theory.

To simplify the solution of the Bethe-Salpeter equation,  the so-called on-shell approximation~\cite{Oller:1997ti,Oset:1997it} is often adopted. It assumes that the interaction kernel can be put on the mass shell with the argument that the off-shell terms can be absorbed by the available coupling constants and physical hadron masses. A vast amount of applications has shown that such an approximation works very well.
Nevertheless, from a formal point of view, one may prefer to take into account the full off-shell effects to have an order-to-order correspondence with the underlying results of chiral perturbation theory. Such off-shell effects have been studied
 for  pion-pion interactions up to next-to-leading order (NLO)~\cite{Nieves:1998hp,Nieves:1999bx} and for
the interactions between the pseudoscalar octet and the ground-state baryon octet up to leading order (LO)~\cite{Nieves:2000km,Nieves:2001wt,Djukanovic:2006xc,Borasoy:2007ku}  and NLO~\cite{Bruns:2010sv,Mai:2012wy}. These studies mainly focused on the description of physical observables such as phase shifts over a wide range of  energies, except in Ref.~\cite{Djukanovic:2006xc} where the contribution to the nucleon mass as a function of the pion mass was studied.

In the present work, we aim to explore whether the off-shell effects in UChPT can lead to an improved description of the light-quark mass dependence of physical observables.\footnote{In the past few years, it has been argued that the light-quark mass dependence of the pole positions of hadronic states may play an important role in revealing their nature. See, e.g., Refs.~\cite{Hanhart:2008mx,Cleven:2010aw}.}  For this purpose, we perform a study of the scattering lengths of Nambu-Goldstone bosons (NGBs) off $D$ mesons~\cite{Liu:2012zya} in  UChPT up to next-to-leading order. To our knowledge, this is the first
of such studies performed in the heavy-light sector, thus extending many previous studies performed with the on-shell approximation~\cite{Guo:2009ct,Liu:2012zya,Wang:2012bu,Altenbuchinger:2013vwa}.  We will show that by taking into account the off-shell terms of the chiral potentials, one can achieve an improved description of the LQCD data~\cite{Liu:2012zya}. On the other hand,
no qualitative difference is observed and therefore our results provide further support to the on-shell approximation for the light-quark mass evolution of  the scattering lengths.

This article is organized as follows. In Sec. II, the relevant chiral  potentials up to NLO are summarized and the formalism to solve the Bethe-Salpeter equation with the
full off-shell dependence is explained.
In Sec. III, we study the latest LQCD simulations of the scattering lengths of  Nambu-Goldstone bosons off $D$ mesons and discuss the implications on the pole position of the dynamically generated $D^*_{s0}(2317)$ resonance.  A short summary is given in Sec. IV.

\section{Theoretical Framework}
\subsection{Chiral potentials up next-to-leading order}
We refer to Ref.~\cite{Altenbuchinger:2013vwa} for the details of the chiral Lagrangians describing the interaction of Nambu-Goldstone bosons and $D$ mesons. As explained there, the $s$ and $u$ channel exchange terms play a negligible role at least in the on-shell approximation. Since our main interest in the present work is to compare the results of the on-shell approach and the off-shell approach, we consider only  the LO (Weinberg Tomozawa) and NLO contact potentials\footnote{It should be noted that
such choices have been adopted as well in Refs.~\cite{Djukanovic:2006xc,Borasoy:2007ku,Bruns:2010sv,Mai:2012wy} since an exact solution of the Bethe-Salpeter equation with general $u$ channel exchange terms has not yet been worked out in UChPT.} of the following form:
\begin{eqnarray}
 \mathcal V_{\rm{WT}}(D(p_1)\phi(p_2)\rightarrow D(p_3)\phi(p_4))&=&
\frac{1}{4 f_0^2}\mathcal C_{\text{LO}}
   \left((p_1+p_2)^2-(p_1-p_4)^2\right)\,,
   \label{LOPot}
   \end{eqnarray}
 \begin{eqnarray}
 \mathcal V_{\rm{NLO}}(D(p_1)\phi(p_2)\rightarrow D(p_3)\phi(p_4))&=&
-\frac{8}{f_0^2} C_{24} \left(c_2\, p_2\cdot p_4-\frac{c_4}{m^2_P}\left( p_1\cdot p_4\, p_2\cdot
   p_3+ p_1\cdot p_2\, p_3\cdot p_4\right)\right)
   \nonumber\\&&
    - \frac{4}{f_0^2} \mathcal C_{35}
   \left(c_3\, p_2\cdot p_4-\frac{c_5}{m^2_P} \left(p_1\cdot p_4 \,p_2\cdot
   p_3+p_1\cdot p_2 \,p_3\cdot p_4\right)\right)\nonumber\\
   &&
   -\frac{8}{f_0^2}
   \mathcal C_0\,c_0
   +\frac{4}{f_0^2} \mathcal C_1\,c_1\, ,
   \label{NLOPot}
\end{eqnarray}
where $p_1(p_3)$ and $p_2(p_4)$ are the four-momenta of the incoming (outgoing) $D$ mesons and Nambu-Goldstone bosons $\phi$, and the coefficients $\mathcal C_{\rm{LO}}$ and $\mathcal C_i$  for different strangeness and isospin combinations $(S,I)$ are listed in Table II of
Ref.~\cite{Altenbuchinger:2013vwa}.

\subsection{Full Bethe-Salpeter equation}
The Bethe-Salpeter equation for a channel of good isospin and strangeness has the following form\footnote{We limit our discussion to a single channel. Extension to coupled channels is straightforward by
promoting $T$, $V$ to matrices.}
\begin{equation}\label{eq:BS}
T(q,Q,P)=V(q,Q,P)+i\int\frac{d^n\tilde{Q}}{(2\pi)^n}V(q,\tilde{Q},P)\frac{1}{(P+\tilde{Q})^2-m^2+i\epsilon}\frac{1}{\tilde{Q}^2-M^2+i\epsilon}T(\tilde{Q},Q,P),
\end{equation}
where $P=p_1+p_2=p_3+p_4$, $q=-p_1$, $Q=-p_3$, and $n$ is the dimension of space-time. To solve the above equation with the kernel $V$ provided by the  chiral potential $V=V_\mathrm{WT}+V_\mathrm{NLO}$, we introduce the following
matrix representation of the potentials:
\begin{equation}
V_{\mathrm{WT}/\mathrm{NLO}}=B(q,P,M_1,\nu,\nu')^T\cdot \hat{V}_{\mathrm{WT}/\mathrm{NLO}}(\nu,\nu',\mu,\mu',P,M_1,M_3)\cdot B(Q,P,M_3,\mu,\mu'),
\end{equation}
\begin{equation}
\hat{V}_\mathrm{WT}=\frac{\mathcal{C}_\mathrm{LO}}{4} \left(
\begin{array}{cccc}
 0 & 0 & 0 & -1\\
 0 & 0 & 0 & 0 \\
 0 & 0 & -2g^{\mu \nu} & 0 \\
 -1& 0 & 0 & \frac{1}{f_0^2}\left(2 P^2-M_1^2-M_3^2\right)
\end{array}
\right),
\end{equation}
\begin{equation}
\hat{V}_\mathrm{NLO}=
f_0^2\left(
\begin{array}{cccc}
 A_1 & 0 & \frac{A_1 P^\mu}{f_0} & \frac{A_1(M_3^2-P^2)}{f_0^2} \\
 0 & A_1g^{\mu \nu'}g^{\mu'\nu} & \frac{ -A_1 P^\nu g^{\mu \nu'}}{f_0} & 0 \\
 \frac{A_1P^\nu}{f_0} & \frac{-A_1 P^\mu g^{\mu'\nu}}{f_0} & \frac{(-8C_{24}c_2-4C_{35}c_3)g^{\mu \nu}+2A_1 P^\mu P^\nu }{f_0^2} & \frac{P^\nu A_1 (M_3^2-P^2)}{f_0^3} \\
 \frac{A_1(M_1^2-P^2)}{f_0^2} & 0 & \frac{P^\mu A_1 (M_1^2-P^2)}{f_0^3} & \frac{-8C_0 c_0+4 C_1c_1 + A_1(P^2-M_1^2)(P^2-M_3^2)}{f_0^4}
\end{array}
\right),
\end{equation}
\begin{equation}
B(q,P,M_1, \nu,\nu')^T=\left(\frac{q^2-M_1^2}{f_0^2},\frac{(P+q)_\nu (P+q)_{\nu'}}{f_0^2},\frac{(P+q)_\nu}{f_0},1\right),
\end{equation}
where $P$ is the center of mass momentum, $M_1$ and $M_3$ are the masses of the initial and final $D$ mesons,  $f_0$ is the pseudoscalar decay constant in the chiral limit, and $A_1=\frac{4}{m_P^2}(2c_4 C_{24}+c_5 C_{35})$. It should be noted that the mapping of the chiral potential into
a matrix form is not unique, but as long as the matrix form allows us to rewrite the BS equation to an  algebraic equation, they are all equivalent.

With the introduction of the matrix representation, the Bethe-Salpeter equation of Eq.~(\ref{eq:BS}) becomes an algebraic equation  of the following form
\begin{equation}\label{eq:BSA}
\hat{T}^{\nu\nu'\mu\mu'}=\hat{V}^{\nu\nu'\mu\mu'}+\hat{V}^{\nu\nu'\rho\rho'} \cdot \hat{G}^{\rho\rho'\sigma\sigma'} \cdot \hat{T}^{\sigma\sigma'\mu\mu'},
\end{equation}
where we have neglected the explicit dependence on $P$, $M_1$, and $M_3$.

The loop function matrix $\hat{G}$ is defined as
\begin{equation}
\hat{G}^{\rho\rho'\sigma\sigma'}=i\int\frac{d^n\tilde{Q}}{(2\pi)^n} \frac{B(\tilde{Q},P,M,\rho,\rho')B(\tilde{Q},P,M,\sigma,\sigma')^T}{[(P+\tilde{Q})^2-m^2+i\epsilon][\tilde{Q}^2-M^2+i\epsilon]},
\end{equation}
and it is now a $4\times4$ matrix.  With the Passarino-Veltman reduction technique, one can easily obtain a representation of $\hat{G}$ in terms of the center of mass momentum $P^\mu$, the metric tensor $g^{\mu\nu}$, and
one-loop scalar 1-point and 2-point functions~\cite{'tHooft:1978xw}.

Upon iterating  the kernel $V_{\mathrm{WT}/\mathrm{NLO}}$, one can identify the most general  solution of Eq.~(\ref{eq:BSA}), $\hat{T}$, to be of the following form

\begin{eqnarray}{}
\footnotesize
&&\hat{T}^{\nu\nu'\mu\mu'}=\nonumber\\
&&\hspace{-0.5cm}\left(
\begin{array}{cccc}
t_{11} & g^{\mu \mu '} t_{12 a}+P^{\mu } P^{\mu '} t_{12 b} &
  P^{\mu } t_{13} & t_{14} \\
g^{\nu \nu '} t_{21 a}+P^{\nu } P^{\nu '} t_{21 b} & P^{\mu }
  P^{\nu } P^{\mu '} P^{\nu '} t_{22 a} +A^{\nu \nu '\mu
  \mu '}+ C^{\nu \nu '\mu \mu' }& P^{\mu } P^{\nu }
  P^{\nu '} t_{23 a}+ B^{\mu \nu \nu' }_{23} & P^{\nu } P^{\nu
  '} t_{24 a}+g^{\nu \nu '} t_{24 b} \\
P^{\nu } t_{31} & P^{\mu } P^{\nu } P^{\mu '} t_{32 a}+
  B^{\nu \mu \mu' }_{32} & g^{\mu \nu } t_{33 a}+P^{\mu } P^{\nu }
  t_{33 b} & P^{\nu } t_{34} \\
t_{41} & P^{\mu } P^{\mu '} t_{42 a}+g^{\mu \mu '} t_{42 b} &
  P^{\mu } t_{43} & t_{44}
\end{array}
\right)\nonumber
\end{eqnarray}
where
\begin{eqnarray*}
C^{\nu\nu'\mu\mu'}&=&g^{\mu \mu'} g^{\nu \nu'}t_{22b}+g^{\nu
  \mu'} g^{\nu'\mu }t_{22c}+g^{\nu \mu } g^{\nu'\mu'}t_{22d},\nonumber\\
B_{23}^{\mu\nu\nu'}&=& P^{\nu'} g^{\nu \mu }t_{23b}+P^{\nu } g^{\nu'\mu }t_{23c}+P^{\mu } g^{\nu \nu'}t_{23d},\nonumber\\
B_{32}^{\nu\mu\mu'}&=& P^{\mu'} g^{\nu \mu }t_{32b}+P^{\mu } g^{\nu\mu' }t_{32c}+P^{\nu } g^{\mu \mu'}t_{32d},\nonumber\\
A^{\nu\nu'\mu\mu'}&=&P^{\mu'} P^{\nu'} g^{\nu \mu }t_{22e}+P^{\nu }P^{\nu'}g^{\mu \mu'}t_{22f}+P^{\nu } P^{\mu'} g^{\nu'\mu }t_{22g}\\
                            &&+P^{\mu } P^{\nu'} g^{\nu \mu' }t_{22h}+P^{\mu}  P^{\mu'}g^{\nu \nu' }t_{22i}+P^{\mu } P^{\nu } g^{\nu' \mu'}t_{22j},
\end{eqnarray*}
and the 36 $t_i$'s are scalar functions of the  squared center-of-mass energy $s=P^2$. A more detailed exposition on how the $t_i$'s are determined is given in the appendix for the LO kernel case.

Several comments on the computation of $\hat{G}$ are in order. Relativistic loop functions involving a heavy particle, whose mass does not vanish in the chiral limit, contain the so-called power-counting breaking (PCB) terms. In the one-baryon sector, various approaches have been proposed to remove the PCB terms, such as the heavy-baryon (HB) formulation~\cite{Jenkins:1990jv}, the infrared (IR) formulation~\cite{Becher:1999he}, and the extended-on-mass shell (EOMS) approach~\cite{Fuchs:2003qc} (see, Ref.~\cite{Geng:2013xn} for a short review about their respective advantages and limitations). Traditionally, in UChPT with the on-shell approximation, no attention is paid to this particular fact since the effects of the PCB terms are effectively absorbed by the so-called subtraction constants (for a recent discussion see Ref.~\cite{Altenbuchinger:2013vwa}). In the studies taking into account the off-shell terms of the chiral potentials, the heavy-baryon formalism is adopted in Refs.~\cite{Nieves:2000km,Nieves:2001wt}, the IR formulation in Ref.~\cite{Djukanovic:2006xc}, and an approach similar in spirit to the EOMS formulation was adopted in Refs.~\cite{Borasoy:2007ku,Bruns:2010sv,Mai:2012wy}. One should note that, however, because of the loss of exact crossing symmetry, in principle one cannot remove the PCB terms by a redefinition of the available low-energy constants (LECs) at the working order in UChPT. Therefore, all the three formulations, the HB, the IR, and the EOMS, should be viewed only as an ansatz to calculate the loop diagrams.

In the present work, in order to compare with the results of the on-shell approximation, we calculate the loop functions $\hat{G}$ in the modified minimal subtraction ($\overline{\mathrm{MS}}$) scheme as in the on-shell approximation~\cite{Altenbuchinger:2013vwa}. Furthermore, we set the renormalization scale at 1 GeV and add one single subtraction constant to the one-loop scalar 1-point and 2-point functions for all the channels, i.e.,
replacing $\log(\mu^2)$ by $\log(\mu^2)+a$. It is well known that in the off-shell scheme the loop functions become more divergent compared to the on-shell loop function. Nevertheless, these loop functions are uniquely regularized in the modified minimal subtraction scheme. Since in the off-shell scheme, an order by order matching to the perturbative ChPT results is possible, one can introduce (at least) a subtraction constant for each scalar function, which can vary between different coupled channels (determined by isospin and strangeness). Then matching to the perturbative results at the appropriate order will allow one to fix or constrain these subtraction constants. J. Nieves and collaborators have studied this in great detail for both the pion-pion and pion-nucleon interactions. See, e.g., Refs.~\cite{Nieves:1998hp,Nieves:1999bx,Nieves:2000km,Nieves:2001wt}. However, such studies are only possible if one has enough experimental data which allow the extra subtraction constants to be fixed. The situation in the present sector does not allow us to perform such a study. As we will see, even with only one common subtraction constant (or renormalization scale) we can already obtain a $\chi^2/\mathrm{d.o.f}$ smaller than 1. Therefore, we will leave such a comprehensive study for a future work once more LQCD/experimental data become available.

\section{Results and discussions}
The scattering lengths of Nambu-Goldstone bosons off $D$ mesons have recently been studied on the lattice~\cite{Liu:2012zya,Mohler:2012na,Mohler:2013rwa} and the $D^*_{s0}(2317)$ is found to be a bound state in the $DK$ channel~\cite{Mohler:2013rwa}.  For our purpose, we focus on the $n_f=2+1$ simulations of Ref.~\cite{Liu:2012zya}, where scattering lengths in five isospin-strangeness channels are obtained at four pion (light-quark) masses with $m_\pi=301,364,511,617$ MeV. They  have obtained the corresponding $D$ and $D_s$ masses as well. Using these heavy-light meson masses, together with those of their physics counterparts, we are able to determine the values of the  LECs $c_0$ and $c_1$ (for details see Ref.~\cite{Altenbuchinger:2013vwa}). As a result, at NLO, we have five unknown LECs to determine, $c_2$, $c_3$, $c_4$, $c_5$, and a subtraction constant $a$, while at LO, only $a$ is unknown. As in Ref.~\cite{Altenbuchinger:2013vwa}, the pseudoscalar decay constant $f_0$ is fixed to that of the pion, 92.21 MeV~\cite{Beringer:1900zz}, unless otherwise stated.

In the present framework, the scattering length of a physical channel  with strangeness $S$ and isospin $I$ is
related to the $T$-matrix element $T$ via
\begin{equation}
a^{(S,I)}=-\frac{1}{8\pi(M_1+m_2)}T^{(S,I)} (s=(M_1+m_2)^2),
\end{equation}
with
\begin{equation}
T^{(S,I)}=B(q,P,M_1,\nu,\nu')^T\cdot \hat{T}^{\nu\nu'\mu\mu'}\cdot B(Q,P,M_3,\mu,\mu')
\end{equation}

\begin{figure}
\includegraphics[width=1.0\textwidth]{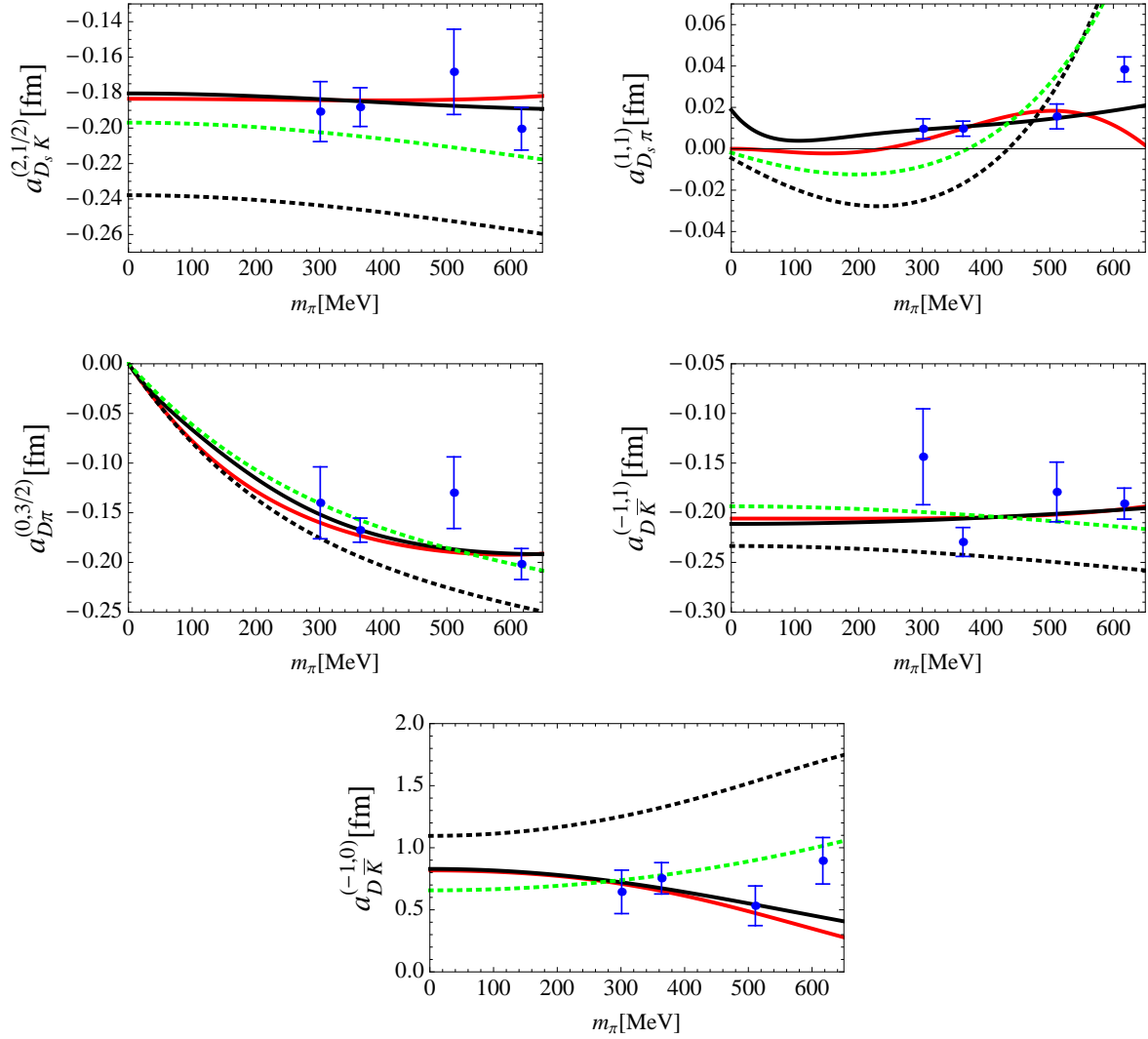}
\caption{The $n_f=2+1$ LQCD data~\cite{Liu:2012zya} vs. the  UChPT fits. The black and red solid lines show
the NLO off-shell and on-shell UChPT fits. The black  and green dashed lines are the LO off-shell UChPT fits with $f_0=92.21$ MeV and $f_0=106.04$ MeV, respectively. \label{fig:lat}}
\end{figure}

\begin{table*}[t]
      \renewcommand{\arraystretch}{1.2}
     \setlength{\tabcolsep}{0.1cm}
     \centering
     \caption{\label{tab:fit2} Low-energy constants, the subtraction constants, and the $\chi^2/\mathrm{d.o.f}$ from the best fits to the LQCD data~\cite{Liu:2012zya} in the
     off-shell UChPT. }
     \begin{tabular}{ccccccc}
     \hline\hline
     & $a$ & $c_{2}$ & $c_{3}$ & $c_4$ & $c_5$ & $\chi^2/\mathrm{d.o.f}$ \\
  LO & $-0.453(11)$ & $$ & $$ & $$ & $$ & 17.9 \\
NLO&0.639(131) & $0.382(181)$ & $0.653(345)$ & $0.597(92)$ & $-2.084 (276)$ & 0.79\\
\hline\hline
    \end{tabular} 
\end{table*}
Fitting these unknown LECs to the lightest 15 LQCD data, we obtain the results shown in Fig.~1, with the corresponding LECs tabulated in
Table I. At leading order, the $\chi^2/\mathrm{d.o.f}\approx17.9$ is rather poor, indicating the failure of a quantitative description of the LQCD data.\footnote{The same is true for the on-shell approach.}
On the other hand, if one would use the SU(3) average of the pseudoscalar decay constants, $f_0=1.15 f_\pi$, instead of $f_\pi$, the $\chi^2/\mathrm{d.o.f}$ would be  reduced to about $3.9$.  At next-to-leading order, we obtain a $\chi^2/\mathrm{d.o.f}\approx0.79$, which should be compared to that obtained in the on-shell approximation, $\chi^2/\mathrm{d.o.f}=1.23$~\cite{Altenbuchinger:2013vwa}.\footnote{It should be mentioned that one could still obtain a $\chi^2/\mathrm{d.o.f.}\approx0.89$ by fitting the whole 20 LQCD data points in the off-shell approach.}  Furthermore, it is interesting to note that the off-shell results and the on-shell results differ most for $a_{D_s\pi}$ in the $(S=1,I=1)$ channel, which couples $D_s\pi $ to $D K$~\cite{Liu:2012zya}.

Clearly, the off-shell effects seem to  improve  the description of the LQCD data of Ref.~\cite{Liu:2012zya}. This result should not be a total surprise. In Refs.\cite{Oller:1997ti,Oset:1997it}, it was pointed out that the off-shell effects, which manifest themselves through diagrams renormalizing the vertices and the hadron masses,  can be absorbed by the available LECs and physical masses. Of course, such ``renormalizations'' are only possible at the physical point, because otherwise one may have to use light-quark mass dependent couplings.\footnote{The hadron masses have to be light-quark mass dependent anyway.} Therefore, to study the light-quark mass evolution of physical observables, one may wish to explicitly keep all the off-shell effects.  Our results  show that this may indeed improve the description of light-quark mass dependence. Nevertheless, more studies are needed in order to confirm that this is generally true.

The LQCD simulations of Ref.~\cite{Liu:2012zya} did not include the ($S=1,I=0)$ channel where the $D^*_{s0}(2317)$ appears. The studies in Refs.~\cite{Liu:2012zya,Altenbuchinger:2013vwa} show that a fit to the LQCD data yields naturally the $D^*_{s0}(2317)$ in UChPT. It will be interesting to check whether this still holds in the present formalism. Searching for a pole in the complex plane, we find a bound state at $\sqrt{s}=2.295$ GeV, which is not so far away from the $D^*_{s0}(2317)$ pole position from the on-shell approach, $\sqrt{s}=2.317$ GeV, which coincides with the experimental measurement. The discrepancy of about 22 MeV provides another  indicator on the magnitude of the off-shell effects.

\begin{figure}
\includegraphics[width=1.0\textwidth]{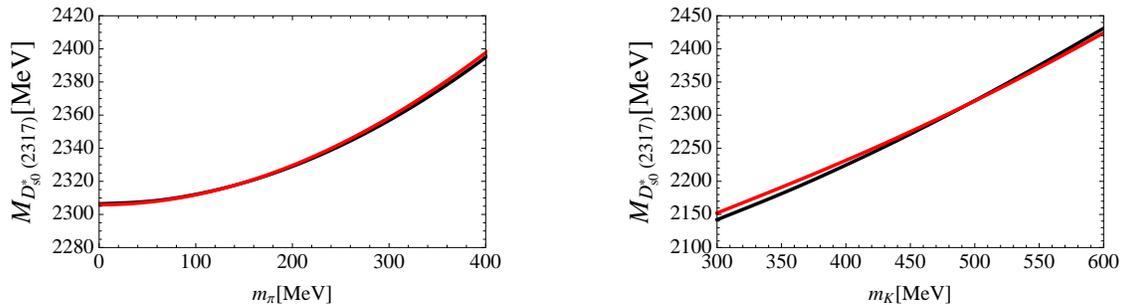}
\caption{Pion (left) and kaon (right) mass evolution of the $D^*_{s0}(2317)$ pole position in $(S=1,I=0)$. The black and red lines are obtained in the off-shell and on-shell UChPT, while the off-shell results have been shifted by 22 MeV to agree with the on-shell results at the physical point. \label{fig:pole}}
\end{figure}

In Fig.~\ref{fig:pole}, we show the pion mass and kaon mass evolution of the $D^*_{s0}(2317)$. To facilitate the comparison, we have shifted the off-shell results  by $22$ MeV so that the on-shell and off-shell results agree at the physical point. The figure in the left panel is obtained by fixing the strange quark mass to its physical value using the leading-order ChPT, while the figure in the right panel is obtained by
fixing the pion mass to its physical value. The dependences of the $D$ and $D_s$ masses on the pion and kaon masses are provided by the next-to-leading-order ChPT as in Ref.~\cite{Altenbuchinger:2013vwa}. It is clear that for the light-quark mass evolution of the $D^*_{s0}(2317)$ pole position
 there is no appreciable difference between the on-shell and off-shell UChPT.
\section{Summary}
We have solved the Bethe-Salpeter equation in unitarized chiral perturbation theory by taking into account the off-shell terms of the chiral potentials up to next-to-leading order. To quantify the magnitude and impact of the off-shell effects, we have studied the latest $n_f=2+1$ LQCD simulations of the scattering lengths of Nambu-Goldstone bosons off $D$ mesons. In comparison with the widely used on-shell approximation, we have shown that taking into account off-shell effects can indeed improve the description of the LQCD data, in terms of light-quark mass evolution. On the other hand,  both descriptions look qualitatively similar, at least for the observables we studied. Therefore, unless the LQCD data become more precise, the on-shell approximation may still be confidently used, given its simplicity.

\section{Acknowledgements}
M. A. and L. S. G. thank Norbert Kaiser and Wolfram Weise for enlightening discussions and a careful reading of the manuscript. This work is supported in part by BMBF, by the A.v. Humboldt foundation, the Fundamental Research Funds for the Central Universities, the National Natural Science Foundation of China (Grant No. 11005007),  the
New Century Excellent Talents in University Program of Ministry of Education of China under
Grant No. NCET-10-0029, the DFG  Cluster of Excellence ``Origin and Structure of the Universe,"
and by DFG and NSFC through the Sino-German CRC 110 ``Symmetries and Emergence of Structure in QCD."

\section{Appendix}

\begin{figure}
\centering
\includegraphics[width=
0.9\textwidth]{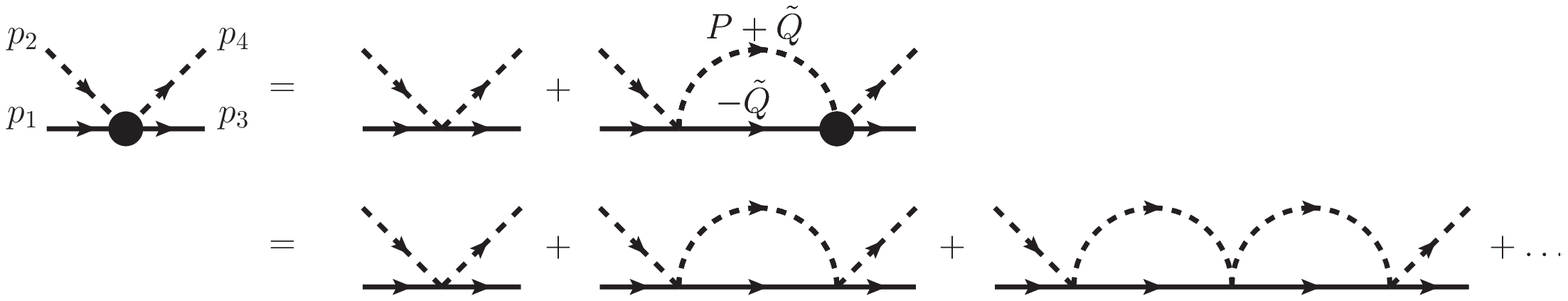}
\caption{Diagrammatical representation of the Bethe-Salpeter equation.
The dashed lines represent Nambu-Goldstone bosons and the solid lines the heavy-light mesons.
The momenta $q$ and $Q$ are related to those of the external mesons $p_1$, $p_2$, $p_3$, and $p_4$ via the following relations: $p_1=-q$, $p_2= q+P$, $p_3= -Q$ and $p_4= Q+P$. }
\label{DiaBS}
\end{figure}
In this appendix we present more details on how the Bethe-Salpeter equation with full off-shell potentials is solved. For the sake of simplicity, we restrict ourselves to the case with the kernel being the leading-order Weinberg-Tomozawa term. In this case, one only needs to introduce a three-component vector $B$,
\begin{equation} \label{BVec}
B(q,P,M,\mu)^T=\left(\frac{q^2-M^2}{f_0^2},\frac{P_{\mu }+q_{\mu }}{f_0},1\right)\,,
\end{equation}
instead of a four-component vector as done in the main text. The Weinberg-Tomazawa potential can then be represented in the
following way:
\begin{eqnarray}
V_{\mathrm{WT}}(q,Q,P)&=&\frac{\mathcal C_{\mathrm{LO}}}{4 f_0^2}(2 P^2-2 (P+q)\cdot (P+Q)-q^2-Q^2)\nonumber\\
&=&B(q,P,M_1,\nu)^T\cdot \hat V_{\mathrm{WT}}(\nu,\mu)\cdot B(Q,P,M_3,\mu)\nonumber\\&\equiv& B(q)^T\cdot \hat V_{\mathrm{WT}}\cdot B(Q)\,,
\label{ItLOMat}
\end{eqnarray}
where
\begin{equation}
\hat V_{\mathrm{WT}}(\nu,\mu)=
\mathcal C_{\textrm{LO}}\left(
\begin{array}{cccc}
 0 & 0 & -\frac{1}{4} \\
 0  & -\frac{g^{\mu \nu }}{2} & 0 \\
 -\frac{1}{4} & 0  & \frac{1}{4f_0^2} \left(2P^2-M_1^2-M_3^2\right)
\end{array}
\right)\,.
\label{LoPotMat}
\end{equation}
In the last line of Eq.~(\ref{ItLOMat}) we have introduced a brief notation such that the summation over Lorentz indices and the dependence on $P$, $M_1$, and $M_3$ are implicit. The idea behind this matrix representation of the potential is that the dependence on the loop momenta, here $q$ and $Q$, can be absorbed into the vector $B$ or the loop matrix $\hat G$ as defined in Eq. (9). In the present case,
the loop matrix becomes a $3\times 3$ matrix (instead of $4\times4$),
\begin{eqnarray}
\hat G(\nu,\mu)&\equiv&\left(
\begin{array}{ccc}
 G_{11}(s)  & 0 & G_{13}(s)\\
 0&g^{\mu\nu} G_{22a}(s)+P^{\mu}P^{\nu} G_{22b}(s) & P^\nu G_{23}(s) \\
 G_{13}(s)  & P^\mu G_{23}(s)&G_{33}(s)
\end{array}
\right)\,.
\label{LoopFunct}
\end{eqnarray}

With these definitions, the leading one-loop diagram of the infinite sum illustrated in Fig.~\ref{DiaBS} simply reads
\begin{eqnarray}
\label{NNLObubble}
&&B(q,P,M_1,\nu)^T\cdot \hat V_{\mathrm{WT}}(\nu,\rho)\cdot \hat G(\rho,\sigma)\cdot \hat V_{\mathrm{WT}}(\sigma,\mu)\cdot B(Q,P,M_3,\mu)\nonumber\\&\equiv& B(q)^T\cdot \hat V_{\mathrm{WT}}\cdot \hat G\cdot \hat V_{\mathrm{WT}}\cdot B(Q)\,,
\end{eqnarray}
and the Bethe-Salpeter equation becomes
\begin{eqnarray}
\label{MatrixBS}
B(q)^T\cdot\hat T\cdot B(Q)&=& B(q)^T\cdot (\hat V_{\mathrm{WT}}+ \hat V_{\mathrm{WT}}\cdot \hat G\cdot \hat V_{\mathrm{WT}}+ \hat V_{\mathrm{WT}}\cdot \hat G\cdot \hat V_{\mathrm{WT}}\cdot \hat G\cdot \hat V_{\mathrm{WT}}+\ldots)\cdot B(Q)\nonumber\\
&=& B(q)^T\cdot (\hat V_{\mathrm{WT}}+ \hat V_{\mathrm{WT}}\cdot \hat G\cdot \hat T)\cdot B(Q)\,.
\end{eqnarray}
 It is clear that $\hat T$ still corresponds to a geometric series, as in the on-shell approximation case. One can easily deduce that $\hat T$ is of the following form
\begin{eqnarray}
&\hat T(\nu,\mu)\equiv\left(\footnotesize{
\begin{array}{cccc}
 t_{11}  &
   P^{\mu } t_{12} & t_{13}\\
 P^{\nu } t_{21}& g^{\mu \nu } t_{22 a}+P^{\mu } P^{\nu }
    t_{22 b} & P^{\nu } t_{23} \\
 t_{31} &
   P^{\mu } t_{32} & t_{33}
\end{array}
}\right)\,,
\label{Ansatz}
\end{eqnarray}
where the functions $t_i$ depend on the involved masses and the total four-momentum squared $s=P^2$. One should notice that Lorentz indices in  $\hat T$ can only be carried by the metric tensor $g^{\mu\nu}$ or the total momentum $P^{\mu}$ ($P^\nu$). More explicitly, for a single-channel process ($M_1=M_3=M$), the Bethe-Salpeter equation
\begin{equation}
\hat T= \hat V_{\mathrm{WT}}+ \hat V_{\mathrm{WT}}\cdot \hat G\cdot \hat T
\end{equation}
gives a set of coupled linear equations,
\begin{eqnarray}
\begin{array}{lll}
 t_{11}&=&-\frac{1}{4} \mathcal{C}_\mathrm{LO} \left(G_{23} P^2 t_{21}+G_{13} t_{11}+G_{33} t_{31}\right) \\
 t_{12}&=&-\frac{1}{4} \mathcal{C}_\mathrm{LO} \left(G_{23} \left(t_{22 a}+P^2 t_{22 b}\right)+G_{13} t_{12}+G_{33} t_{32}\right)\\
 t_{13}&=&-\frac{1}{4} \mathcal{C}_\mathrm{LO} \left(G_{23} P^2 t_{23}+G_{13} t_{13}+G_{33} t_{33}+1\right) \\
 t_{21}&=&-\frac{1}{2} \mathcal{C}_\mathrm{LO} \left(t_{21} G_{22 a}+P^2 t_{21} G_{22 b}+G_{23} t_{31}\right)\\
 t_{22a}&=&-\frac{1}{2} \mathcal{C}_\mathrm{LO} \left(G_{22 a} t_{22 a}+1\right)\\
 t_{22b}&=&-\frac{1}{2} \mathcal{C}_\mathrm{LO}  \left(G_{22 b} \left(t_{22 a}+P^2 t_{22 b}\right)+G_{22 a} t_{22 b}+G_{23} t_{32}\right) \\
 t_{23}&=&-\frac{1}{2} \mathcal{C}_\mathrm{LO} \left(t_{23} G_{22 a}+P^2 t_{23} G_{22 b}+G_{23} t_{33}\right)\\
 t_{31}&=&-\frac{1}{4} \mathcal{C}_\mathrm{LO} \left(-2 \Delta  G_{23} P^2 t_{21}-2 \Delta  G_{33} t_{31}+G_{13} \left(t_{31}-2 \Delta  t_{11}\right)+G_{11} t_{11}+1\right)\\
 t_{32}&=&-\frac{1}{4} \mathcal{C}_\mathrm{LO} \left(-2 \Delta  \left(G_{23} \left(t_{22 a}+P^2 t_{22 b}\right)+G_{33} t_{32}\right)+G_{13} \left(t_{32}-2 \Delta  t_{12}\right)+G_{11} t_{12}\right) \\
 t_{33}&=&\frac{1}{4} \mathcal{C}_\mathrm{LO} \left(2 \Delta  \left(G_{23} P^2 t_{23}+G_{33} t_{33}+1\right)+G_{13} \left(2 \Delta  t_{13}-t_{33}\right)-G_{11} t_{13}\right)\,,\nonumber
\end{array}
\end{eqnarray}
with $\Delta=(P^2-M^2)/f_0^2$.  These equations determine all the $t_i$'s introduced in Eq.~(\ref{Ansatz}). The generalization to coupled channels is straightforward.

\end{document}